\documentclass[aps,pra,graphicx,superscriptaddress,twocolumn]{revtex4-1}

\usepackage{graphicx}
\usepackage{hyperref}
\usepackage{epstopdf}
\usepackage{bbm}
\usepackage{xcolor}
\usepackage{ulem}
\usepackage{subfigure}
\usepackage{amsmath}
\usepackage{amssymb}

\newcommand{\bra}[1]{\langle #1 |}
\newcommand{\ket}[1]{| #1 \rangle}

\begin{document}

\title{Engineering optical hybrid entanglement\\between discrete- and continuous-variable states}

\author{Kun Huang}
\email{khuang@usst.edu.cn}
\affiliation{Shanghai Key Laboratory of Modern Optical Systems, and Engineering Research Center of Optical Instruments and Systems (Ministry of Education), School of Optical Electrical and Computer Engineering, University of Shanghai for Science and Technology, Shanghai 200093, China}

\author{Hanna Le Jeannic}
\affiliation{Niels Bohr Institute, University of Copenhagen, Blegdamsvej 17, DK-2100 Copenhagen, Denmark}

\author{Olivier Morin}
\affiliation{Max-Planck-Institut für Quantenoptik, Hans-Kopfermann-Strasse 1, D-85748 Garching, Germany}

\author{Tom Darras}
\affiliation{Laboratoire Kastler Brossel, Sorbonne Universit\'e, CNRS, ENS-Universit\'e PSL, Coll\`ege de France, 4 Place Jussieu, 75005 Paris, France}

\author{Giovanni Guccione}
\affiliation{Laboratoire Kastler Brossel, Sorbonne Universit\'e, CNRS, ENS-Universit\'e PSL, Coll\`ege de France, 4 Place Jussieu, 75005 Paris, France}

\author{Adrien Cavaill\`{e}s}
\affiliation{Laboratoire Kastler Brossel, Sorbonne Universit\'e, CNRS, ENS-Universit\'e PSL, Coll\`ege de France, 4 Place Jussieu, 75005 Paris, France}

\author{Julien Laurat}
\email{julien.laurat@sorbonne-universite.fr}
\affiliation{Laboratoire Kastler Brossel, Sorbonne Universit\'e, CNRS, ENS-Universit\'e PSL, Coll\`ege de France, 4 Place Jussieu, 75005 Paris, France}

\date{\today}

\begin{abstract}
The generation and manipulation of hybrid entanglement of light involving discrete- and continuous-variable states have recently appeared as essential resources towards the realization of heterogeneous quantum networks. Here we investigate a scheme for the remote generation of hybrid entanglement between particle-like and wave-like optical qubits based on a non-local heralding photon detection. We also extend this scheme with additional local or non-local detections. An additional local heralding allows the resulting state to exhibit a higher fidelity with the targeted entangled qubits while a two-photon non-local heralding detection gives access to a higher dimensionality in the discrete-variable subspace, resulting thereby in the generation of hybrid entangled qutrits. The implementation of the presented schemes, in combination with ongoing works on high-fidelity quantum state engineering, will provide novel non-classical light sources for the development of optical hybrid architectures.
\end{abstract}

\maketitle

\section{Introduction}
Following the wave-particle duality of light, optical quantum information protocols have been traditionally implemented based either on discrete variables or on continuous variables of the electromagnetic field \cite{OBrien2009NP,Furusawa2011Book,roadmap2018}. The discrete-variable (DV) approach involves single photons living in a finite-dimensional space spanned for instance by orthogonal polarizations or the absence or presence of a single photon \cite{Kok2007RMP}. Alternatively, the continuous-variable (CV) approach encodes information within quadrature components (amplitude or phase) of a light field \cite{Braunstein2005RMP} and qubits can be implemented as arbitrary superpositions of coherent states with equal amplitude but opposite phases, also known as optical Schr\"{o}dinger cat states or qumodes \cite{Ralph2003PRA,Ourjoumtsev2006Science,Loock2011LPR}.

Recently, tremendous progress has been seen to combine both approaches in a so-called optical hybrid architecture, with the aim of gathering benefits from both sides and exploring new capabilities in quantum information science \cite{Loock2011LPR,Son2002JMO,Paternostro2004PRL, Andersen2015NPh}. First of all, DV operations are necessary to transform initial Gaussian resources to non-Gaussian states \cite{Eisert2002PRL}. Photon subtraction acting on squeezed vacuum states is now commonly used to prepare Fock states and their superpositions, including Schr\"{o}dinger cat states \cite{Ourjoumtsev2006Science,Ourjoumtsev2006PRL, Yukawa2013OE, Bimbard2010NP, Morin2012OL, Huang2016PRA, Jeannic2016OL, Polzik2006,Takahashi2008PRL, Morin2013PRL1, Morin2014JOVE, Huang2015PRL,Jeannic2018PRL}. Beyond quantum state engineering, the DV-CV hybridization has also led to a variety of theoretical studies and experimental investigations of novel protocols where states and operations are combined. Experimental implementations of Gaussian entanglement distillation have been made possible by coherent subtraction of single photons \cite{Ourjoumtsev2007PRL, Takahashi2010NP, Ulanov2015NP}. Deterministic continuous-variable teleportation of DV qubits has been demonstrated \cite{Takeda2013} as well as single-photon entanglement certification based on local quadrature measurements between distant nodes \cite{Morin2013PRL2, Ho2014NJP}.

Another motivation in the field is the development of heterogeneous quantum networks, where CV and DV resources can be effectively combined and interconverted \cite{Loock2011LPR, Andersen2015NPh}. In this context, the ability to interface disparate encoding basis becomes a key component for quantum teleportation protocols. In this scenario the required quantum link is provided by hybrid entanglement of light, i.e., entanglement of the form $\ket{0}\ket{\alpha}+\ket{1}\ket{-\alpha}$ between particle-like and wave-like optical qubits \cite{Kreis2012PRA}. Recently such hybrid entangled states have been experimentally generated \cite{Morin2014NP, Jeong2014NP,Costanzo2015}, including at a distance via a loss-tolerant scheme \cite{Morin2014NP}. These entangled states have then been used to interconvert quantum information between CV and DV encodings \cite{Ulanov2017PRL, Sychev2018NC}, to demonstrate remote state preparation of arbitrary CV qubits by local manipulation of the DV component \cite{Jeannic2018Optica} or to violate a steering inequality that shows their suitability for semi-device-independent protocols \cite{Cavailles2018}. Such class of states has also been considered for resource-efficient quantum computation \cite{Qi2015CPB, Loock2008PRA} and near-deterministic quantum teleportation \cite{Lee2013PRA, Park2012PRA}. Additionally, optical hybrid entanglement is reminiscent of the original idea of the Schr\"{o}dinger cat state, with potential interesting studies of micro-macro properties in phase space \cite{Andersen2013PRA, Frowis2018RMP}.

The hybrid entangled state first generated in \cite{Morin2014NP} relies on a probabilistic preparation heralded by the detection of a single photon in an indistinguishable fashion. Here we provide a detailed investigation of this scheme and extend it to engineer CV-DV entangled states. In particular, we investigate the effect of local squeezing and photon-counting balancing, and we consider various cases with local and non-local photon subtractions. The presented schemes will also stimulate fundamental exploration on properties of such class of states, which may be the basis for novel protocols and practical implementation of heterogeneous quantum networks.

This paper is organized as follows. First, in section \ref{sec:hybrid} we outline a measurement-induced scheme for the preparation of hybrid qubit entanglement and illustrate a possible application of the generated state as a quantum encoding converter. Then, in section \ref{sec:hybrid1} we provide a formal derivation of the state and discuss the experimental steps for its preparation as realized in \cite{Morin2014NP}. We evaluate in particular the properties of the entangled states and discuss the decoherence induced by photon loss or phase noise. In section \ref{sec:hybrid1add} we investigate an enhanced scheme, where an additional local subtraction is applied to improve the fidelity with the targeted state. Then, in section \ref{sec:hybrid2} we introduce the non-local detection of two photons to generate hybrid qutrit entanglement for higher dimensionality. The effects of local squeezing and photon-counting balancing are discussed. Finally, in section \ref{sec:conclusion} we provide a short summary and outlook.

\section{Measurement-induced generation of hybrid entanglement}
\label{sec:hybrid}

In principle, deterministic generation of hybrid entanglement between discrete- and continuous-variable can be realized by dispersive light-matter interaction \cite{vanloock2006} or by cross-Kerr nonlinearity between a coherent state and a single-photon qubit \cite{Gerry1999PRA, Jeong2005PRA}. However, such approaches are still very challenging. A compromised workaround resorts to the seminal Knill-Laflamme-Milburn protocol \cite{Knill2001Nature}, where the desired strong nonlinear coupling is provided by conditional photon detection at the expense of a probabilistic operation. In this section we consider the scheme for the generation of remote CV-DV hybrid entanglement of light based on the heralding of a photon at a central station. This scheme will be the starting point for all ensuing discussions.

\begin{figure}[tb!]
\centering
\includegraphics[width=0.48\textwidth]{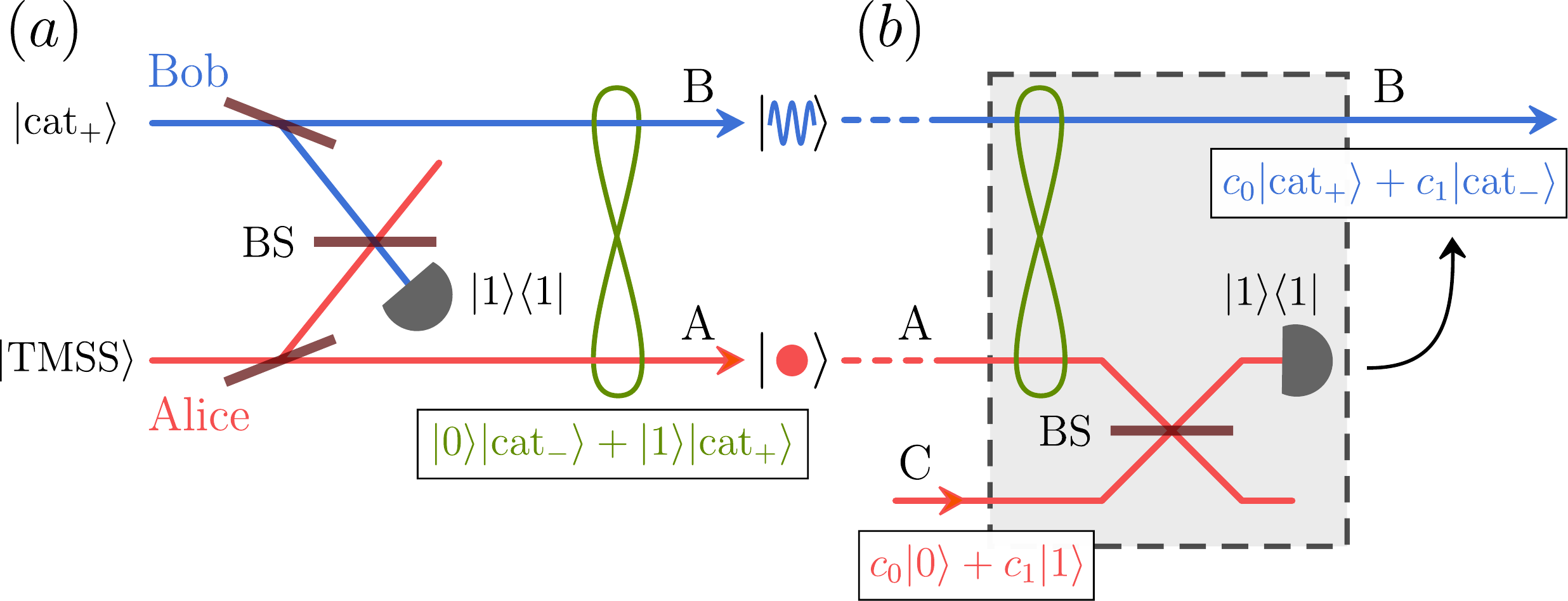}
\caption{Hybrid entanglement of light used as the resource for a qubit converter. (a) Generation scheme for hybrid entanglement between particle-like and wave-like optical qubits. The small fraction subtracted from an even cat state $\ket{\textrm{cat}_+}$ on Bob's side is mixed with one mode of a weak two-mode squeezed vacuum state $|\textrm{TMSS}\rangle$ generated on Alice's side. A single-photon detection at the central station will herald the generation of hybrid entanglement. (b) Scheme for the qubit encoding converter. The key resource for this teleportation-based protocol is the hybrid entanglement of light. The input qubit is mixed with the discrete-variable mode of the hybrid entangled state on a symmetric beam-splitter. The detection of a single photon in one output port of the beam-splitter will herald the successful mapping to the targeted continuous-variable qubit.}
\label{fig:hybrid}
\end{figure}

The scheme is illustrated in figure \ref{fig:hybrid}a. The basis for discrete encoding is provided by the absence and presence of a single photon, $\ket{0}$ and $\ket{1}$. For the continuous-variable qubit, the basis is given by even and odd coherent state superpositions, also known as cat states, $\ket{\textrm{cat}_+}$ and $\ket{\textrm{cat}_-}$. A small fraction of light is subtracted from Bob's initial state, $\ket{\textrm{cat}_+}$, and mixed with one mode of a weak two-mode squeezed vacuum $\left|\textrm{TMSS}\right\rangle$ generated on Alice's side. A single-photon detection at one output of the beam-splitter will herald the generation of hybrid entanglement. The resulting output state, for which a rigorous derivation is provided in section \ref{sec:generalmodel}, is of the form $\ket{0}\ket{\textrm{cat}_-}+\ket{1}\ket{\textrm{cat}_+}$.

One advantage of this scheme is that the fragile components remain local, and only single photons propagate between the two distant nodes. As a result, a lossy heralding channel affects the success rate but not the fidelity of the resulting state, such as in the Duan-Lukin-Cirac-Zoller proposal for entangling remote quantum memories \cite{DLCZ}. Therefore, the proposed method is suitable to establish hybrid entanglement connection over a long distance.

\subsection{An example application: a qubit encoding converter}

One example of a networking protocol based on hybrid entanglement is a qubit encoding converter, which would allow the interaction and transfer of information between different types of qubits in heterogeneous quantum networks.

A possible converter scheme is presented in figure \ref{fig:hybrid}b, where the hybrid entangled state of light serves as the driving element of a discrete-to-continuous converter mapping single-photon qubits to coherent-state-superposition qubits. The converter, represented by the gray box, receives an arbitrary superposition between presence and absence of a single photon as input, $c_0\ket{0}_C+c_1\ket{1}_C$, which is then mixed on a beam-splitter with the DV mode of the hybrid entangled state $\ket{0}_A\ket{\textrm{cat}_-}_B+\ket{1}_A\ket{\textrm{cat}_+}_B$. The successful conversion is heralded by the detection of \textit{exactly} one photon at one output port of the beam-splitter within the converter. This single photon can originate either from mode $A$ or mode $C$. If the single photon comes from the input qubit state, then the heralded state will be $c_1\ket{\textrm{cat}_-}_B$. If instead the single photon is coming from the hybrid state, the resulting output will be $c_0\ket{\textrm{cat}_+}_B$. Due to the indistinguishability between these two events, a qubit state is obtained as $c_0\ket{\textrm{cat}_+}_B + c_1\ket{\textrm{cat}_-}_B$.

Rigorously, the mixing between the input qubit and the hybrid entangled state on the symmetric beam-splitter are described by the following operator evolutions:
\begin{equation}
\hat{a}^\dagger \to \frac{\hat{a}^\dagger + \hat{c}^\dagger}{\sqrt{2}}\quad \textrm{and} \quad \hat{c}^\dagger \to \frac{\hat{c}^\dagger - \hat{a}^\dagger}{\sqrt{2}}\ ,
\end{equation}
where $\hat{a}^\dagger$ and $\hat{c}^\dagger$ are creation operators in modes $A$ and $C$. The resulting three-mode state is given by
\begin{equation}
\begin{split}
\ket{\psi} &\propto c_0\ket{0}_C\ket{0}_A\ket{\textrm{cat}_-}_B+c_1(\ket{2}_C\ket{0}_A-\ket{0}_C\ket{2}_A)\ket{\textrm{cat}_+}_B\\
&\quad+\frac{c_0}{\sqrt{2}}(\ket{1}_C\ket{0}_A+\ket{0}_C\ket{1}_A)\ket{\textrm{cat}_+}_B\\
&\quad+\frac{c_1}{\sqrt{2}}(\ket{1}_C\ket{0}_A-\ket{0}_C\ket{1}_A)\ket{\textrm{cat}_-}_B\ .
\end{split}
\end{equation}
Projecting on a single photon on mode $C$ and tracing out mode $A$, the result is the targeted qumode on mode $B$ converted from the original input. Note that the reverse conversion from qumodes to qubits can be realized by using Bell-state measurements for coherent-state qubits \cite{Jeong2001PRA}. In these converters, hybrid entanglement is the critical resource for mapping the two computational bases. 

\section{Generation of hybrid qubit entanglement}
\label{sec:hybrid1}

In this section we focus on the creation of hybrid entanglement between particle-like and wave-like optical qubits. We consider current experimental generation schemes, as demonstrated in \cite{Morin2014NP}, and identify the relevant parameters and resources. In particular, we discuss the negativity of the associated Wigner functions and the negativity of entanglement of the bipartite state, that will also be considered in the extended schemes presented in the next sections.

\subsection{Model of the generation scheme}
\label{sec:generalmodel}

\begin{figure}[tt!]
\centerline{
\includegraphics[width=0.45\textwidth]{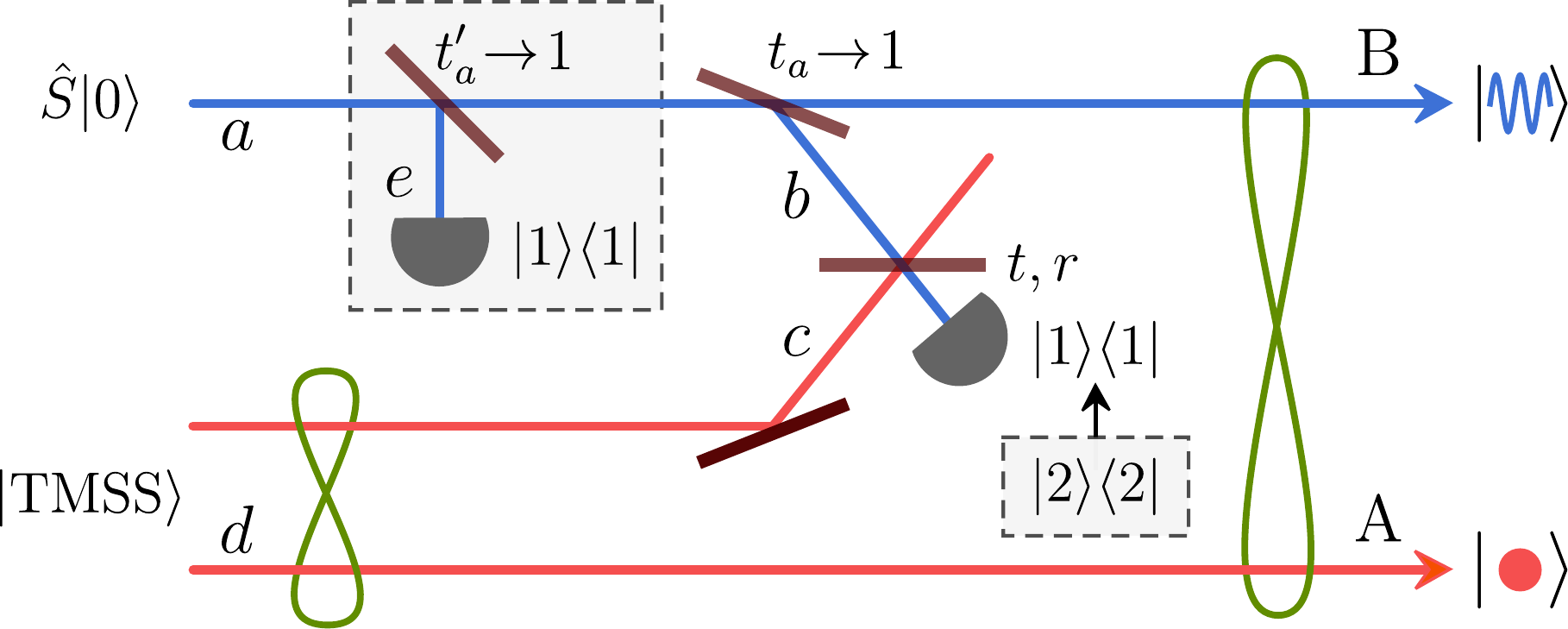}}	
\caption{Scheme for measurement-induced generation of hybrid entanglement in a practical scenario. The initial continuous-variable state, an even cat state $\ket{\textrm{cat}_+}$, is approximated here by a squeezed vacuum, as experimentally realized in \cite{Morin2014NP}. The optional shaded boxes represent extensions of the scheme. The first box on mode $a$ enables the implementation of an additional local subtraction on Bob's side to increase the fidelity and size of the generated hybrid entangled states, as discussed in section \ref{sec:hybrid1add}. The second optional box replaces the measurement at the central station with a two-photon detection, enabling the generation of hybrid qutrit entanglement as detailed in section \ref{sec:hybrid2}.}
	\label{fig:hybridscheme}
\end{figure}

A detailed generation scheme is illustrated in figure \ref{fig:hybridscheme}. On the continuous-variable side, the Schr\"{o}dinger cat states can be expressed as
\begin{equation}
\ket{\textrm{cat}_\pm}= \frac{\ket{\alpha}\pm\ket{-\alpha}}{N_\pm} = \frac{\ket{\alpha}\pm\ket{-\alpha}}{\sqrt{2(1\pm e^{-2\alpha^2})}}\ ,
\end{equation}
where the amplitude $\alpha$ of the coherent state is assumed to be real for simplicity. A small fraction of light is tapped from the initial cat state by a beam-splitter of amplitude reflectivity $\sin(\theta) \approx \theta \ll 1$, resulting in
\begin{equation}
\begin{split}
\hat B(\theta)\ket{\textrm{cat}_+}_a \ket{0}_b &= \text{e}^{\theta (\hat a \hat b^\dag - \hat a^\dag\hat b)}\ket{\textrm{cat}_+}_a \ket{0}_b \\
&\approx (1 + \theta \hat a \hat b^\dag) \ket{\textrm{cat}_+}_a \ket{0}_b\ ,
\end{split}
\end{equation}
where $\hat B(\theta)$ is the beam-splitting operator and $\hat a^\dag$, $\hat b^\dag$ are the respective creation operators for modes $a$ and $b$. On the other side a two-mode squeezed vacuum state in the low gain limit (with a small squeezing factor $\lambda \ll 1$) is prepared, which can be approximated as
\begin{equation}
\begin{split}
\ket{\textrm{TMSS}}_{c,d} &\approx \ket{0}_c\ket{0}_d+ \lambda \ket{1}_c\ket{1}_d \\
&= (1 + \lambda \hat c^\dag \hat d^\dag)\ket{0}_c\ket{0}_d\ ,
\end{split}
\end{equation}
where $\hat c^\dagger$ and $\hat d^\dagger$ are creation operators in the corresponding modes $c$ and $d$.

The modes $b$ and $c$ are then spatially combined on a beam-splitter with reflectivity and transmissivity coefficients $r$ and $t$, leading to the transformations $\hat b^\dag \to \text{e}^{i\varphi_1}(t\hat b^\dag + r\hat c^\dag)$ and $\hat c^\dag \to \text{e}^{i\varphi_2}(t\hat c^\dag - r\hat b^\dag)$, where $\varphi_{1,2}$ denote the accumulated phases. Substitution within the above expressions yields the resulting state:
\begin{equation}
\begin{split}
(1 + \text{e}^{i\varphi_1}\theta r \hat a \hat c^\dag + \text{e}^{i\varphi_1}\theta t \hat a \hat b^\dag)\times \\
(1 + \text{e}^{i\varphi_2}\lambda t \hat c^\dag \hat d^\dag - \text{e}^{i\varphi_2}\lambda r \hat b^\dag \hat d^\dag) \ket{\textrm{cat}_+}_a\ket{0}_{b,c,d}\ .
\end{split}
\end{equation}
We implement the measurement-induced generation of the entangled state by considering only the terms containing $b^\dag$, i.e.\ ignoring vacuum contributions and considering only heralded states on mode $b$. Tracing out mode $c$ and keeping only first-order terms in $\theta$ and $\lambda$, the state is finally reduced to
\begin{equation}
\label{eq:hybrid1PS}
\theta t\hat a \ket{\textrm{cat}_+}_a\ket{0}_d + \lambda r \ket{\textrm{cat}_+}_a\ket{1}_d \ ,
\end{equation}
with $\Delta \varphi = \varphi_1 - \varphi_2 = \pi$ for simplicity. Since $\hat{a}\ket{\textrm{cat}_+} = \alpha \frac{N_-}{N_+} \ket{\textrm{cat}_-}$, the heralded state can be rewritten as
\begin{equation}
\theta t \alpha N_- \ket{0}_d\ket{\textrm{cat}_-}_a + \lambda r N_+ \ket{1}_d\ket{\textrm{cat}_+}_a\ .
\end{equation}
The superposition weights can be balanced by adjusting the beam-splitter ratio to obtain the targeted maximally hybrid entangled state:
\begin{equation}
\label{targetedState}
\ket{\Psi}_{AB} = \frac{\ket{0}_A\ket{\textrm{cat}_-}_B + \ket{1}_A\ket{\textrm{cat}_+}_B}{\sqrt{2}} \ .
\end{equation}
Note that the subscripts $d$ and $a$ have been replaced by $A$ and $B$ to indicate that the resulting state is shared by Alice and Bob, who use DV and CV encodings, respectively. The entangled state above can alternatively be expressed in the rotated qubit basis, $\ket{+}=(\ket{0}+\ket{1})/\sqrt{2}$ and $\ket{-}=(\ket{0}-\ket{1})/\sqrt{2}$, as
\begin{equation}
\begin{split}
\ket{\Psi}_{AB} = \frac{1}{\sqrt{2}} \Big(\ket{+}_A \frac{\ket{\textrm{cat}_-}_B + \ket{\textrm{cat}_+}_B}{\sqrt{2}} + \\
\ket{-}_A \frac{\ket{\textrm{cat}_-}_B - \ket{\textrm{cat}_+}_B}{\sqrt{2}} \Big)\ ,
\end{split}
\end{equation}
which in the limit of a large $\alpha$ is equivalent to
\begin{equation}
\ket{\Psi}_{AB} = \frac{\ket{+}_A\ket{\alpha}_B - \ket{-}_A\ket{-\alpha}_B}{\sqrt{2}} \ .
\end{equation}
The state can be converted into other forms by the application of targeted operations. A Hadamard gate to the DV mode, which can be realized with a non-Gaussian ancilla and projective measurements \cite{Marek2010PRA}, would for example transform the above hybrid entangled state into $\ket{0}_A\ket{\alpha}_B + \ket{1}_A\ket{-\alpha}_B$. In the following discussion we will commonly refer to expression \ref{targetedState} unless otherwise specified.

\subsection{Experimental resources}
The protocol above shows how hybrid entanglement can be established between two CV and DV optical modes. One element that it does not take into account however is how to create the CV source state. Large-amplitude cat states are generally hard to realize in the optical domain. In the experiment reported in \cite{Morin2014NP} the CV mode is approximated by a squeezed vacuum state $\hat{S}(\zeta)\ket{0}$, which has high fidelity with an even cat state $\ket{\textrm{cat}_+}$ of amplitude $\alpha \lesssim 1$ \cite{Nielsen2007PRA}.

Introducing the squeezed vacuum state and the single-photon subtracted squeezed vacuum state,
\begin{equation}
\label{eq:Sqsub1}
\begin{split}
\ket{\textrm{0PS}} &= \hat{S}(\zeta) \ket{0} \quad \\
\quad \ket{\textrm{1PS}} &= \frac{\hat{a} \hat{S}(\zeta) \ket{0}}{\sinh \zeta} = \hat{S}(\zeta) \ket{1}\ ,
\end{split}
\end{equation}
we can rewrite expression \ref{eq:hybrid1PS} as
\begin{equation}
\theta t \sinh\zeta \ket{0}_A \ket{\textrm{1PS}}_B + \lambda r \ket{1}_A \ket{\textrm{0PS}}_B\ .
\end{equation}
The resulting two-mode state can then be written as
\begin{equation}
\label{eq:hybrid1}
\ket{\Psi}_{AB} = \hat{S}_B(\zeta) \frac{\ket{0, 1}_{A,B} + \mu \ket{1, 0}_{A,B}}{\sqrt{1+\mu^2}} \ .
\end{equation}
The weight parameter $\mu$ is defined by
\begin{equation}
\mu = \frac{\lambda r}{\theta t \sinh\zeta} = \sqrt {\frac{N_A}{N_B}}
\end{equation}
and it can be extracted directly from the experiment by measuring the photon counts $N_A$ and $N_B$ coming from the two modes taken independently. Symmetric balancing between the two components of the entangled state in a lossless scenario is simply expressed by the condition $\mu^2 = 1$, or equivalently $N_A = N_B$. In the presence of loss, modeled by finite transmission through a beam-splitter, balancing is achieved when $\mu^2 = \eta_B/\eta_A$, where $\eta_A$ and $\eta_B$ are the intensity transmissions for the corresponding channel.

It should be noted that in this low-amplitude regime the \textit{continuous} structure of Bob's mode is determined by the squeezing operation $\hat{S}_B(\zeta)$. Given the local nature of this operation, we will bypass it whenever appropriate in order to operate with a rather simplified model to obtain analytical expressions.

\subsection{Representation of the hybrid entangled state}
\begin{figure}[t]
\centering
\includegraphics[width=1 \columnwidth]{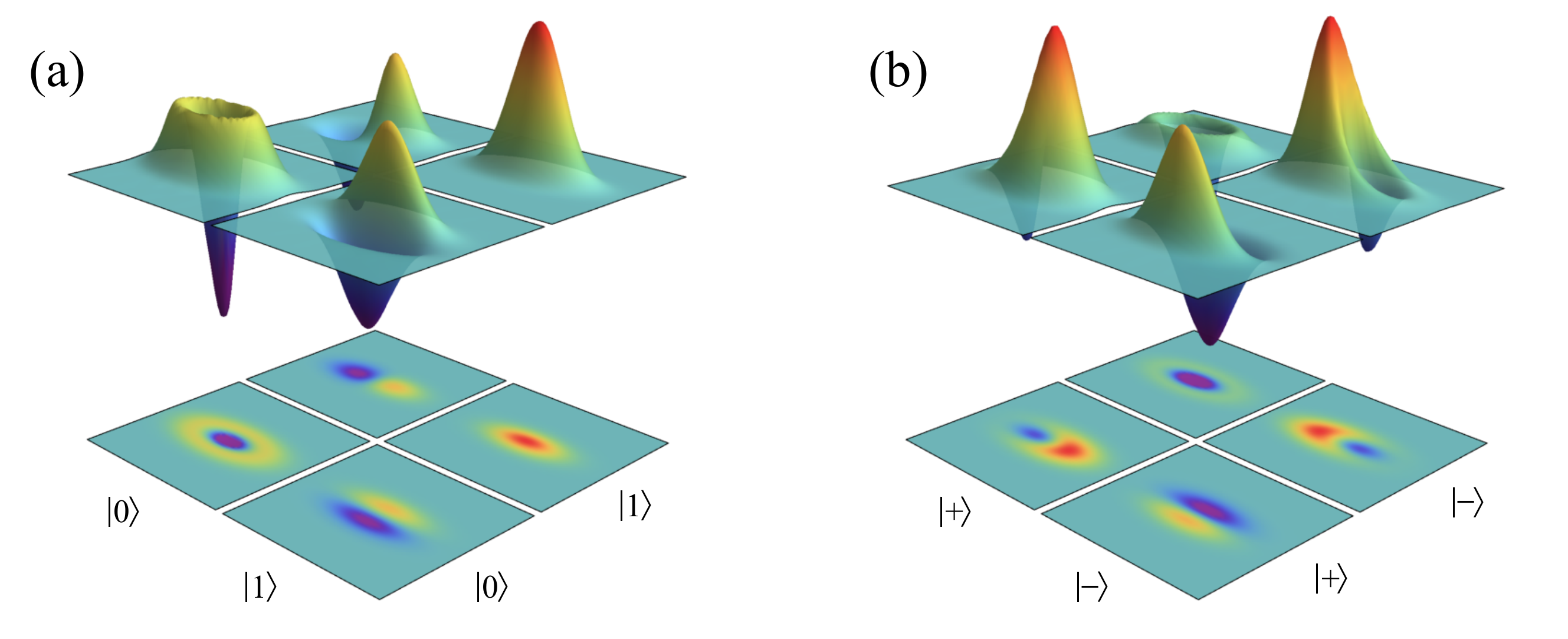}
\caption{Hybrid entangled state representation. (a) The blocks provide the Wigner functions associated with the reduced density matrices $\bra{k}\hat{\rho}\ket{l}$ with $k,l$ $\in$ $\{0,1\}$. Since the components with $k\neq l$ are not hermitian, the corresponding Wigner functions are not necessarily real, but conjugate. The plot shows the real part for $k>l$ and the imaginary part for $k<l$. (b) Wigner functions associated with the reduced density matrices $\langle k |\hat{\rho}| l \rangle$ with $k,l$ $\in$ $\{+,-\}$, where $|+\rangle$ and $|-\rangle$ stand respectively for the rotated basis $(|0\rangle+|1\rangle)/\sqrt{2}$ and $(|0\rangle-|1\rangle)/\sqrt{2}$. Here we consider symmetric balancing with $\mu^2 = 1$ and no losses. The local squeezing on Bob's side is 3 dB.}
\label{fig:fig3}
\end{figure}

\begin{figure*}[t]
\centerline{
\includegraphics[width=0.80\textwidth]{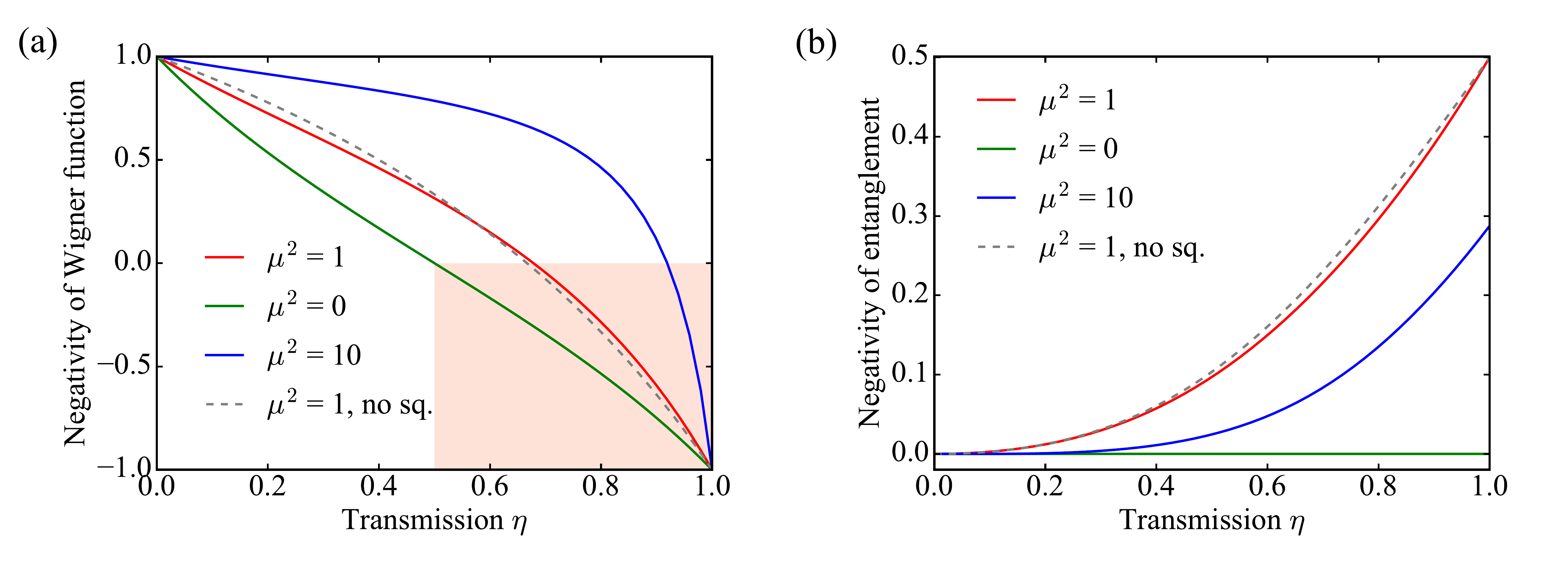}}	
\caption{(a) Negativity of the Wigner function of the state $\bra{0}\hat \rho _{A,B}\ket{0}$ as a function of the intensity transmission for various balancing factors. The losses in two modes are assumed to be symmetric, with $\eta_A=\eta_B=\eta$. The shaded region indicates the condition to obtain a negative Wigner function. (b) Negativity of entanglement for various balancing factors as a function of the transmission $\eta$. The local squeezing on Bob's side is 3 dB. The dashed lines in gray correspond to the simplified model by removing the local squeezing.}
\label{fig:fig4}
\end{figure*}

Typically, the DV quantum states living in a finite-dimensional Hilbert space are described in terms of density matrices, while the CV states with potentially infinite dimensionality are displayed in the Wigner function representation. Here we combine the two methods to show the hybrid entangled state in a visual manner. Specifically, the Wigner functions are used to represent the reduced density matrices $\bra{k}\hat\rho_{A,B}\ket{l}$, where $\ket{k}$ and $\ket{l}$ indicate states in the discrete basis. The hybrid representation of the two-mode density matrix is shown in figure \ref{fig:fig3}(a) for the case of symmetric balancing $\mu^2 = 1$ and no losses. The diagonal blocks $\bra{0}\hat{\rho}\ket{0}$ and $\bra{1}\hat{\rho}\ket{1}$ correspond to the single-photon subtracted squeezed vacuum state and the squeezed vacuum state, respectively. The non-zero off-diagonal terms show the coherence of the superposition. Note that the Wigner functions associated with these off-diagonal terms are complex conjugate. The real and imaginary parts are thus used for the terms $\bra{1}\hat\rho_{A,B}\ket{0}$ and $\bra{0}\hat\rho_{A,B}\ket{1}$, respectively.

The generated hybrid state can also be represented in the rotated DV basis formed by $\{\ket{+}=(\ket{0}+\ket{1})/\sqrt{2},\ \ket{-}=(\ket{0}-\ket{1})/\sqrt{2}\}$, as shown in figure \ref{fig:fig3}(b). The two projections $\bra{+}\hat\rho_{A,B}\ket{+}$ and $\bra{-}\hat\rho_{A,B}\ket{-}$ exhibit an opposite displacement in phase space. In this basis, the diagonal terms resemble coherent states $\ket{\alpha}$ and $\ket{-\alpha}$ with opposite phases. However, they deviate from the round Gaussian profile expected for coherent states, showing in particular negative values. This feature arises because of the initial approximation that uses a squeezed vacuum as the initial source. As a consequence, the maximum achievable fidelity with the targeted hybrid state of size $|\alpha|^2 = 1$ given in equation \ref{targetedState} is limited to 92\%. To go beyond this limitation, Bob can perform a local single-photon subtraction to initially prepare an odd cat sate. The details on this enhanced scheme will be given in section \ref{sec:hybrid1add}.

\subsection{Negativity of the Wigner function}
As can be seen in figure \ref{fig:fig3}(a), the block $\bra{0}\hat \rho_{A,B}\ket{0}$ shows negative values in its Wigner function representation, which is a strong, directly accessible signature of the non-Gaussianity of the hybrid state. In the protocol, the term $\bra{0}\hat \rho_{A,B} \ket{0}$ corresponds to a photon-subtracted squeezed vacuum state heralded by the detection of a single photon on Bob's side. This photon detection is a non-Gaussian operation that leads to a pronounced negativity in the Wigner function, which can be quantitatively evaluated by the value at the origin of phase space.

Figure \ref{fig:fig4}(a) shows this Wigner negativity as a function of symmetric transmission efficiency $\eta_A=\eta_B=\eta$ under different balancing conditions. Note that the Wigner function has been normalized by the scaling factor $2\pi\sigma_0 ^2$, where $\sigma_0 ^2$ is the quadrature variance defined by the vacuum state. In the limiting case $\mu^2=0$ all the heralding photons come from Bob's side and the two-mode state is separable. For a small amount of squeezing, the projected state $\hat S_B(\zeta)\ket{1}_{B}$ is approximated to a single-photon state, whose decoherence follows a linear dependence on transmission as $1 - 2 \eta$. At the other limit, i.e., for a large value of $\mu^2$, the heralding photons come predominantly from Alice's side. After losses the term $\mu\ket{1,0}_{A,B}$ will lead to an appreciable weight of $\ket{0,0}_{A,B}$ compared to that of $\ket{0,1}_{A,B}$. Hence, the dominating vacuum state after projection on the discrete mode $\ket{0}_A\bra{0}$ could decrease the negativity even in the presence of small losses, indicating a faster decoherence for larger values of $\mu^2$ as shown in Fig.\ \ref{fig:fig4}(a). For symmetric balancing $\mu^2 = 1$, the negativity of the Wigner function is present for an intensity transmission above 67.8\%.

Rigorously, the losses in the two modes both affect the negativity \cite{Morin2014NP}. To elaborate this point we consider the approximated state obtained by omitting the small local squeezing on Bob's side, leading essentially to a superposition state of the form $\ket{0,1}_{A,B} + \mu\ket{1,0}_{A,B}$ as shown in equation \ref{eq:hybrid1}. In this case, the evolution of the Wigner function negativity in a lossy channel can be expressed analytically as
\begin{equation}
\mathcal{W} = \frac{(1-2\eta_B)+(1-\eta_A)\mu^2}{1+(1-\eta_A)\mu^2}\ .
\end{equation}
Balancing the state in the general case of asymmetric losses, $\mu^ 2 = \eta_B/\eta_A $, we obtain
\begin{equation}
\mathcal{W} = \frac{\eta_A + \eta_B - 3 \eta_A \eta_B}{\eta_A + \eta_B - \eta_A \eta_B} \ .
\end{equation}
The boundary for negativity can be derived as $1/\eta_A + 1/\eta_B \le 3$. A finite efficiency on Alice's side results into extra vacuum contributions on Bob's side. In the particular case of symmetric channels, the transmission for both modes should be larger than 2/3 to preserve negativity in the Wigner function. This requirement is experimentally challenging but feasible for current high-fidelity quantum state engineering experiments \cite{Yukawa2013OE, Morin2014JOVE}. The obtained boundary agrees well with the previously obtained value of 67.8\%, justifying thereby the simplified model in the limit of small local squeezing. The reduction of the Wigner function negativity given by the simplified model with a 3-dB squeezing on Bob's side is also shown in Fig. \ref{fig:fig4}(a) and exhibits a slight deviation with the full model as expected. 

We note that in the presented simulations, the state preparation losses (initial squeezing resource that can be not pure), the channel propagation losses and the detection losses are incorporated into a total effective loss. Indeed, only the overall loss affects the properties of the heralded state. We also underline that losses in the conditioning path are only decreasing the generation rate and not the fidelity of the heralded state (assuming dark counts contribution to be negligible, which is usually the case for experiments based for instance on superconducting single-photon detectors as used in our implementations).

\subsection{Negativity of entanglement}
The generated entanglement of the bipartite state can be quantitatively measured by computing the entanglement negativity \cite{Vidal2002PRA} defined as
\begin{equation}
\mathcal{N}(\hat{\rho})=\frac{1}{2}\sum_i (|\lambda_i|-\lambda_i),
\end{equation}
where $\lambda_i$ are the eigenvalues of the partial transpose $\rho^{\rm T_A}$. The negativity of entanglement has an upper bound of 0.5, corresponding to a maximally entangled qubit state.

Figure \ref{fig:fig4}(b) gives the negativity of entanglement as a function of the symmetric intensity transmission $\eta$ for various balancing factors. For unbalanced cases with $\mu^2 \neq 1$, the negativity of entanglement is smaller than the maximum value of 0.5, as expected. Particularly, in the limiting cases $\mu^2 = 0$ and $\mu^2 \gg 1$ the obtained hybrid states are separable. For balanced heralding with $\mu^2=1$, the negativity of entanglement decreases with decreasing transmission from the maximum value of 0.5 to 0. Specifically, for the minimum efficiency of 2/3 required to obtain negativity of the Wigner function, the corresponding negativity of entanglement is about 0.2. With the help of advanced optical quantum technologies, which can demonstrate a high efficiency above 90\% in both modes, limited for instance by current escape efficiencies of optical parametric oscillators \cite{Morin2014JOVE}, it is possible to achieve an entanglement negativity $\mathcal{N}=0.4$.

For a small squeezing of the initial source on Bob's side, the simplified model without local squeezing can be used to obtain analytical expressions. In this case, the two-mode density matrix corresponding to the state $\ket{0,1}_{A,B} + \mu\ket{1,0}_{A,B}$ leads to
\begin{equation}
\mathcal{N} = \frac{\sqrt{4\eta^2\mu^2+(1-\eta)^2(1+\mu^2)^2} - (1-\eta)(1+\mu^2)}{2(1+\mu^2)}\ .
\end{equation}
For a squeezing of 3 dB on Bob's side, the decay of entanglement given by this expression is shown in Fig. \ref{fig:fig4}(b), which is closed to the one obtained by the numerical computation based on the full model.

\subsection{Phase noise}
To generate entanglement, the relative phase $\phi$ in the hybrid state $\ket{\Psi}_{AB} \propto \ket{0}_A\ket{\textrm{cat}_-}_B+\text{e}^{i\phi}\ket{1}_A\ket{\textrm{cat}_+}_B$ has to be kept constant over successive heralding events. In any practical realization, however, the superposition phase between the two heralding paths can vary due to imperfect phase locking. Assuming that the phase noise follows a Gaussian distribution with standard deviation $\sigma$, the resulting density matrix then reads
\begin{equation}
\hat{\rho} = \frac{1}{\sqrt{2\pi}\sigma}\int_{-\infty}^{\infty}\text{e}^{-\frac{\phi^2}{2\sigma^2}}\ket{\Psi}_{AB}\bra{\Psi}\text{d} \phi\ .
\end{equation}
Even in the absence of loss ($\eta=1$) the negativity of entanglement will be affected by the phase noise and follow a Gaussian decay $\frac{1}{2}\text{e}^{-\sigma^2/2}$. It is worth noting that the accumulated phase fluctuation is not only due to the instability of the two conditional paths but also to all the perturbations of the other upstream locking signals, such as in the preparation of the initial squeezing source. 

In a typical experiment, the phase parameter could be controlled with a standard deviation smaller than 5$^\circ$ \cite{Morin2014NP}, thereby barely degrading the entanglement negativity. Specifically, for a 5\% drop from the maximum entanglement value (0.5), the allowed standard deviation $\sigma$ would be about 18$^\circ$. Hence phase noise plays here a negligible role compared to other loss factors in the experiment.

In the next section, we will present an enhanced version of the preparation scheme to obtain hybrid entangled states with larger size and better fidelity to the targeted state.

\section{Generation of enhanced hybrid qubit entanglement}
\label{sec:hybrid1add}

In this section we will consider a variation of the previous scheme. A local single-photon subtraction is now implemented on Bob's side to prepare an odd cat state as the initial CV source. After a single-photon heralding event, the entangled state $\ket{0}_A\ket{\textrm{cat}_+}_B+\ket{1}_A\ket{\textrm{cat}_-}_B$ is obtained. From an experimental perspective, the even cat is now approximated by a two-photon subtracted squeezed vacuum, which allows a better approximation of cat states with larger sizes \cite{Nielsen2007PRA}. In the following, we present the enhanced scheme and detail the properties of the heralded state.

\subsection{Model of the generation scheme}
As depicted within the shaded box in figure \ref{fig:hybridscheme}, the additional local photon subtraction is implemented on the initial even cat state on Bob's side. The subsequent operations are identical to the ones used in the aforementioned scheme. This novel implementation thus combines local and non-local photon detections.

Considering an extra mode $e$ for the additional single-photon subtraction, the state on Bob's side evolves as
\begin{equation}
\begin{split}
 \hat B_{ab}(\theta)\hat B_{ae}(\theta_0)\ket{\textrm{cat}_+}_a \ket{0}_b \ket{0}_e \\
 \approx (1+\theta \hat a \hat b^\dag)(1+\theta_0 \hat a \hat e^\dag)\ket{\textrm{cat}_+}_a \ket{0}_b \ket{0}_e\ .
\end{split}
\end{equation}
On Alice's side, the state is still given by the two-mode squeezed vacuum state $\ket{\textrm{TMSS}}_{c,d}$. As before, the conditional paths in modes $b$ and $c$ are combined on a beam-splitter at the central station, leading to the state
\begin{equation}
\begin{split}
[1+\text{e}^{i\varphi_1}\theta \hat a (t\hat b^\dagger+r\hat c^\dagger)](1+\theta_0\hat a \hat e^\dagger) \times\\
[1+\text{e}^{i\varphi_2}\lambda \hat d^\dagger (t\hat c^\dagger-r\hat b^\dagger)]\ket{\textrm{cat}_+}_a \ket{0}_{b,c,d,e}\ .
\end{split}
\end{equation}
Keeping only terms containing $\hat b^\dag$ and $\hat e^\dag$ of first order in $\theta$ and $\lambda$, one obtains
\begin{equation}
(e^{i\varphi_1}\theta t\hat a^2 \hat b^\dag \hat e^\dag - e^{i\varphi_2}\lambda r \hat a \hat b^\dag \hat d^\dag \hat e^\dag)\ket{\textrm{cat}_+}_a \ket{0}_{b,c,d,e}\ .
\end{equation}
By projecting modes $b$ and $e$ onto the single-photon state and tracing out mode $c$, the state becomes
\begin{equation}
\theta t \hat a^2 \ket{0}_A \ket{\textrm{cat}_+}_B+ \lambda r\hat a \ket{1}_A \ket{\textrm{cat}_-}_B\ ,
\end{equation}
where $\Delta \varphi = {\varphi _1} - {\varphi _2} = \pi$ is used and subscripts $d, a$ are replaced by $A,B$.

In an experimental realization, the initial even cat state can be approximated by a squeezed vacuum state. Using again the states of equation \ref{eq:Sqsub1} and introducing the two-photon subtracted squeezed vacuum state,
\begin{equation}
\label{eq:Sqsub2}
\begin{split}
\ket{\textrm{2PS}} &= \frac{\hat{a}^2 \hat{S}(\zeta) \ket{0}}{\sinh \zeta \sqrt{1+3\sinh^2 \zeta}} \\
&=  \frac{\cosh \zeta \hat{S}(\zeta)\ket{0} + \sqrt{2} \sinh \zeta \hat{S}(\zeta)\ket{2}}{\sqrt{1+3\sinh^2 \zeta}} \ ,
\end{split}
\end{equation}
we can finally rewrite the entangled state as
\begin{equation}
\label{eq:hybridAddSub}
\ket{\Psi}_{AB} = \sqrt{3+1/\sinh^2\zeta}\ket{0}_A\ket{\textrm{2PS}}_B + \mu \ket{1}_A \ket{\textrm{1PS}}_B\ .
\end{equation}

\subsection{Photon-counting balancing condition}
To obtain maximal entanglement for the hybrid state given in equation \ref{eq:hybridAddSub} the relative weights should be equalized as
\begin{equation}
\label{eq:hybrid2add_condition}
\mu^2 = 3 + \frac{1}{\sinh^2\zeta} = 2(1+c^2)\ ,
\end{equation}
where we use $c = 1/(\sqrt 2 \tanh \zeta)$. As before, this condition can be experimentally achieved by adjusting the beam-splitter ratio at the central station for mixing the two conditional paths. For the sake of clarity, we will refer to this as the \textit{two-photon} balancing condition in order to distinguish it from the \textit{single-photon} balancing condition, $\mu^2=1$, previously obtained.

We can interpret the two-photon balancing condition from the perspective of photon counting. We assume that $N_0$ is the photon count from the local subtraction on Bob's side, and $N_A$ and $N_B$ are the photon counts in the conditional path from Alice's and Bob's side, without subtraction, for an acquisition time $T$. Therefore, to maximize the indistinguishability we need to balance the two coincidence counts:
\begin{equation}
\begin{split}
 C_{0,A} &= g_A N_0 N_A \tau/T  \\
 C_{0,B} &= g_B N_0 N_B \tau/T \ ,
 \end{split}
\end{equation}
where $g_A$ and $g_B$ are the degrees of correlation for the corresponding coincident detections and $\tau$ the time width of the coincidence window. Since the photon counts $N_0$ and $N_B$ are from the same squeezed vacuum source, the auto-correlation function is given by $g_B=3+1/\sinh^2\zeta$ . In contrast, the counts $N_0$ and $N_A$ are from two uncorrelated sources, leading to $g_A=1$. Therefore, balancing the coincidence counts between the two modes requires
\begin{equation}
\begin{split}
C_{0,A} =C_{0,B} &\Rightarrow \frac{g_B}{g_A} = \frac{N_A}{N_B} \\
&\Rightarrow 3 + \frac{1}{\sinh^2\zeta} = \mu^2\ ,
 \end{split}
\end{equation}
which corresponds to the condition given by equation \ref{eq:hybrid2add_condition}.

\subsection{Representation of the hybrid entangled state}

\begin{figure}[t]
\includegraphics[width=1\columnwidth]{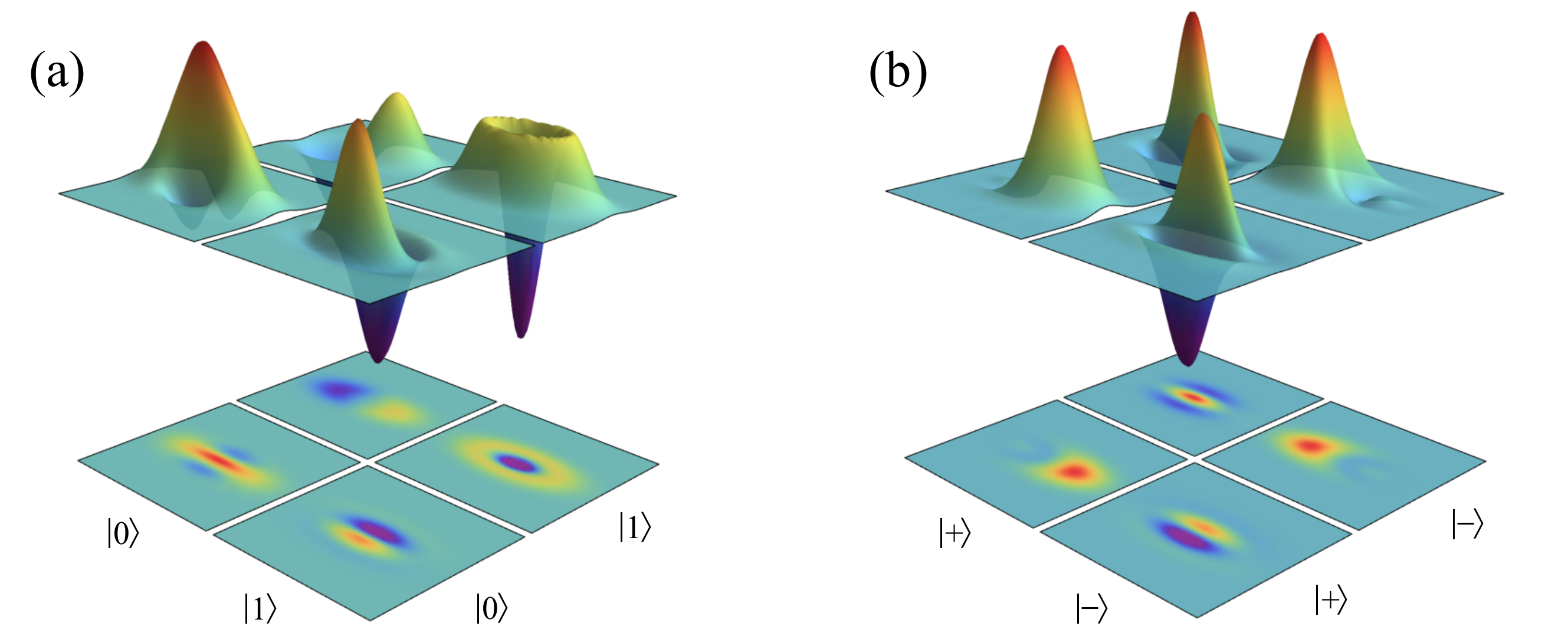}
\centering
\caption{Hybrid representation of the heralded hybrid entangled state with an additional local subtraction in the lossless, symmetrically-balanced scenario. The blocks provide the Wigner functions associated with the reduced density matrices $\bra{k}\hat{\rho}\ket{l}$ with $k,l \in$ $\{0,1\}$ in (a) and $k,l \in$ $\{+,-\}$ in (b), respectively. The local squeezing on Bob's side is 3 dB.}
\label{fig:fig5}
\end{figure}

Figure \ref{fig:fig5}(a) displays the generated hybrid entangled state in the case of an initial 3-dB squeezed vacuum. The diagonal blocks $\bra{0}\hat{\rho}_{A,B}\ket{0}$ and $\bra{1}\hat{\rho}_{A,B}\ket{1}$ correspond to the single-photon-subtracted squeezed vacuum and the two-photon-subtracted squeezed vacuum, respectively. Additionally, the hybrid state in the rotated DV basis is shown in figure \ref{fig:fig5}(b). The projected states in the diagonal are mostly positive with a fairly round shape, i.e. a better similarity to coherent states compared to the case without using the local subtraction as shown in figure \ref{fig:fig3}(b). This feature results from the use of a photon-subtracted squeezed vacuum state as the initial source for the entangled state. A quantitative comparison of the fidelities to the targeted hybrid states in the two scenari will be given later in section \ref{sec:HybridFidelity}.

\subsection{Negativity of the Wigner function}

We now evaluate the negativity of the Wigner function corresponding to the state $\bra{1}\hat\rho_{A,B}\ket{1}$. As given in equation \ref{eq:hybridAddSub}, the resulting state is a single-photon subtracted vacuum state $\ket{\textrm{1PS}}_B$, which is equivalent to a squeezed single-photon state $\hat S_B(\zeta)\ket{1}_{B}$. The trend of the negativity of the Wigner function in the case of an initial 3-dB local squeezing is given by the green line in Fig. \ref{fig:fig4}(a). For a small amount of squeezing, it decreases linearly with transmission efficiency as
\begin{equation}
\mathcal{W} = 1 - 2 \eta_B\ ,
\end{equation}
which is independent from the efficiency in Alice's mode, the balancing parameter and the squeezing parameter. Indeed, all these parameters affect the preparation rate of the state but not its purity.

\subsection{Negativity of entanglement}
For the proposed enhanced version, the targeted entangled state is still a hybrid qubit state similar to the one produced by the original scheme, albeit with a basis flip in one mode. Hence, the achievable maximum entanglement is expected to be the same value of 0.50. To elaborate on this point, the state presented in \ref{eq:hybridAddSub} can be rewritten as
\begin{equation}
\ket{\Psi}_{AB} = \hat{S}_B(\zeta) \frac{\ket{0}_A(c\ket{0}_B + \ket{2}_B)+\mu/\sqrt 2 \ket{1}_A \ket{1}_B}{\sqrt{1+c^2+\mu^2/2}}\ .
\label{eq:hybrid1flip}
\end{equation}
In the absence of losses, the negativity of entanglement can be derived from this expression:
\begin{equation}
\mathcal{N} = \frac{\mu \sqrt{2+2c^2}}{2+2c^2+\mu^2}.
\end{equation}
A maximum negativity of 0.5 can always be obtained when the two-photon balancing condition, $\mu^2 = 2(1+c^2)$, is satisfied. In this case, the generated state corresponds to a maximally entangled qubit state. 

\begin{figure*}[t]
\centering
\includegraphics[width=0.85\textwidth]{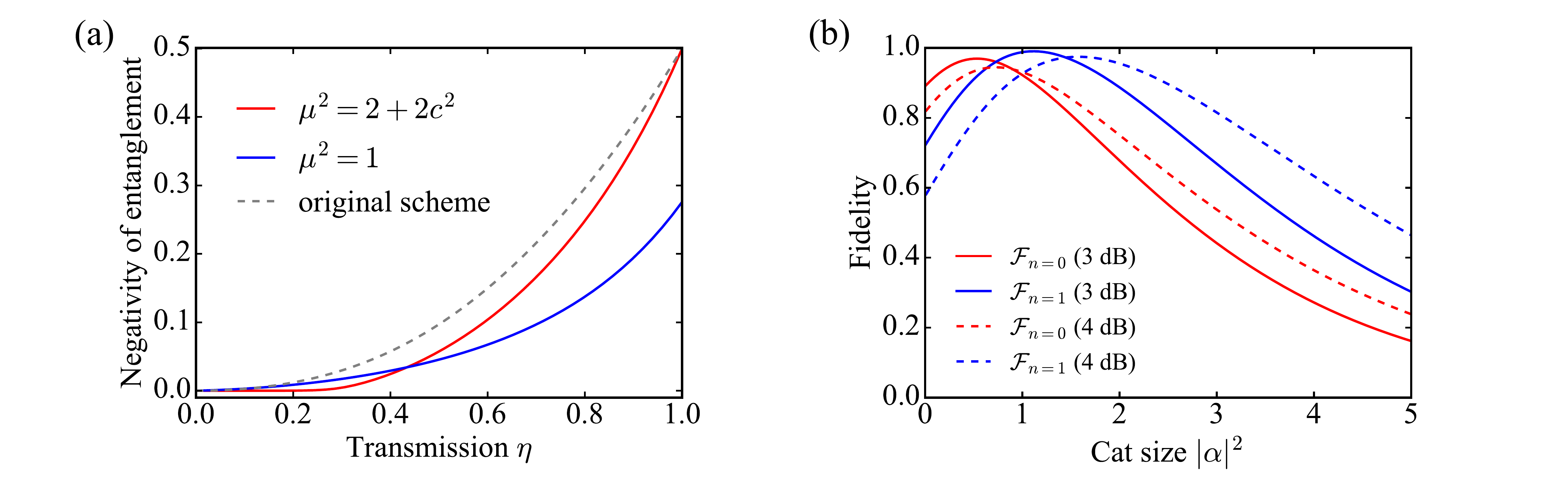}
\caption{(a) Negativity of entanglement as a function of the intensity transmission for the single-photon and two-photon balancing conditions. For comparison, the negativity for the state generated by the original scheme is given with the dashed line. The losses in the two modes are assumed to be symmetric. (b) Fidelities $\mathcal{F}_{n=1}$ and $\mathcal{F}_{n=0}$, i.e.\ with and without local single-photon subtraction on Bob's side, as a function of $|\alpha|^2$ and for different squeezing levels. The implementation of the additional photon subtraction allows to obtain $|\alpha|^2$ above 1 while keeping a high fidelity.}
\label{fig:fig6}
\end{figure*}

Figure \ref{fig:fig6}(a) shows the negativity of entanglement in the presence of loss. The two-photon balancing condition is found to be critical to obtain maximum negativity of entanglement. Under the single-photon balancing condition, i.e. $\mu^2=1$, the negativity is always smaller than 0.28. Additionally, we can observe that the hybrid qubit state prepared by the enhanced scheme is more sensitive to the losses due to the higher photon number components in the CV mode.

\subsection{Fidelity}
\label{sec:HybridFidelity}
Here we compare the two schemes for the generation of hybrid entangled qubits, i.e.\ with and without the implementation of the local photon subtraction on Bob's side. Intuitively, the second scheme should produce hybrid entangled states with higher fidelity since the even cat state in this case is better approximated by a two-photon-subtracted squeezed vacuum state. In the following, we quantify this difference.

In the first scheme, without local subtraction, the maximally entangled hybrid state reads
\begin{equation}
\ket{\Phi_0} = \frac{\ket{0}_A\ket{\textrm{1PS}}_B + \ket{1}_A\ket{\textrm{0PS}}_B}{\sqrt 2}\ ,
\end{equation}
and the targeted state is
\begin{equation}
\ket{\Psi_0} = \frac{\ket{0}_A\ket{\textrm{cat}_-}_B + \ket{1}_A\ket{\textrm{cat}_+}_B}{\sqrt 2}\ .
\end{equation}
The corresponding fidelity for no local subtraction, i.e.\ $n=0$, is obtained as
\begin{equation}
\begin{split}
\mathcal{F}_{n=0} = |\bra{\Psi_0}\Phi_0\rangle|^2 &=\frac{|\bra{cat_+}\textrm{0PS}\rangle + \bra{cat_-}\textrm{1PS}\rangle|^2}{4} \\
&= \frac{\big(\sqrt{\mathcal{F}_0}+\sqrt{\mathcal{F}_1}\big)^2}{4}\ ,
\end{split}
\end{equation}
where $\mathcal{F}_0$ and $\mathcal{F}_1$ are given by
\begin{equation}
\mathcal{F}_0=|\bra{cat_+}\textrm{0PS}\rangle|^2 =\frac{\sqrt{1-\lambda^2}e^{\lambda\alpha^2}}{\cosh\alpha^2}
\end{equation}
and
\begin{equation}
\mathcal{F}_1=|\bra{cat_-}{\textrm{1PS}}\rangle|^2 =\frac{(1-\lambda^2)^{3/2}\alpha^2e^{\lambda\alpha^2}}{\sinh\alpha^2}\ .
\end{equation}
In the second scheme, the maximally entangled hybrid qubit obtained after a local subtraction is
\begin{equation}
\ket{\Phi_1} = \frac{\ket{0}_A\ket{\textrm{2PS}}_B + \ket{1}_A\ket{\textrm{1PS}}_B}{\sqrt{2}}\ .
\end{equation}
The targeted hybrid entangled state is given by
\begin{equation}
\ket{\Psi_1} = \frac{\ket{0}_A\ket{\textrm{cat}_+}_B + \ket{1}_A\ket{\textrm{cat}_-}_B}{\sqrt{2}}\ .
\end{equation}
The fidelity in this case with local single-photon subtraction, i.e.\ $n=1$, is obtained as
\begin{equation}
\begin{split}
\mathcal{F}_{n=1} &= |\bra{\Psi_1}\Phi_1\rangle|^2 =\frac{|\bra{cat_+}\textrm{2PS}\rangle + \bra{cat_-}\textrm{1PS}\rangle|^2}{4} \\
&= \frac{\big(\sqrt{\mathcal{F}_2}+\sqrt{\mathcal{F}_1}\big)^2}{4}\ ,
\end{split}
\end{equation}
where $\mathcal{F}_2$ is given by
\begin{equation}
\mathcal{F}_2=|\bra{cat_+}{\textrm{2PS}}\rangle|^2 =\frac{(1-\lambda^2)^{5/2}(1+\lambda\alpha^2)^2e^{\lambda\alpha^2}}{(1+2\lambda^2)\cosh\alpha^2}\ .
\end{equation}

These fidelities are plotted in figure \ref{fig:fig6}(b) as a function of the cat size $|\alpha|^2$ for different values of the initial squeezing. We can see that the fidelity with the targeted hybrid state is indeed improved by the additional local subtraction to generate cat states with sizes $|\alpha|^2 \gtrsim 1$. Specifically, in the case of 3 dB of squeezing, we have $\mathcal{F}_{n=0} \approx 92\%$ and $\mathcal{F}_{n=1} \approx 99\%$ for $|\alpha|^2 = 1$. By increasing the squeezing, the state retains a high fidelity even for larger cat sizes. For instance, by using 4-dB squeezing, the scheme with local photon subtraction can achieve a cat size of $|\alpha|^2 = 2$ with fidelity of 96\% while the first scheme can only achieve a fidelity around 75\%.

While the enhanced scheme can substantially improve the fidelity of the generated hybrid qubit state, the amount of entanglement is still inherently limited by the dimensionality of the state. In the following section, we propose a modified scheme to generate hybrid qutrit entanglement.

\section{Generation of hybrid qutrit entanglement}
\label{sec:hybrid2}

The scheme shown in figure \ref{fig:hybridscheme} can be extended to generate hybrid qutrit entanglement. For this purpose, a two-photon heralding detection at the central station should be used. The resulting entangled state occupies a higher dimensional Hilbert space, with the discrete mode spanning the $\{\ket{0}, \ket{1}, \ket{2}\}$ subspace. In this section we derive the generated state and detail the related properties.

\begin{figure*}[t]
\centering
\includegraphics[width=0.8 \textwidth]{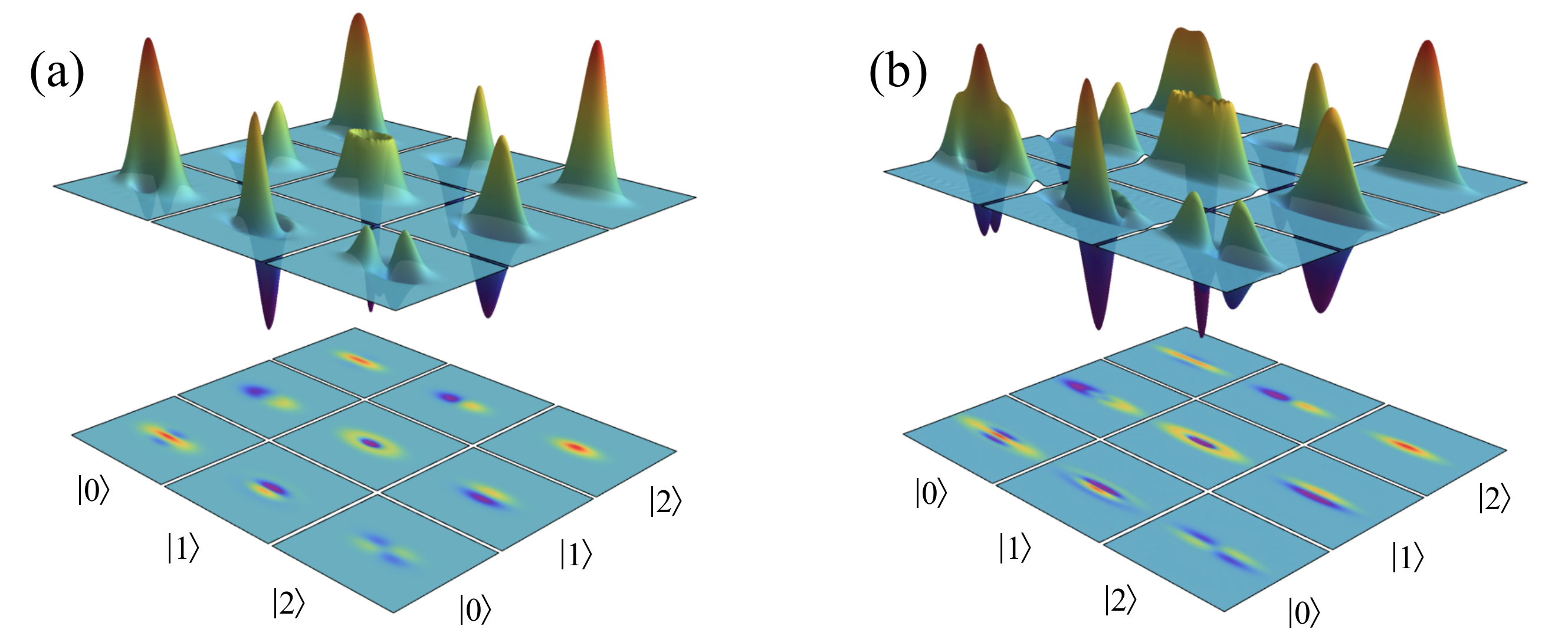}
\caption{Hybrid entangled qutrit states under the balancing condition for maximal entanglement $\mu^4 = 1 + c^2$, for a local squeezing of 3 dB (a) and 6 dB (b), respectively. The diagonal blocks $\bra{0}\hat \rho_{A,B}\ket{0}$, $\bra{1}\hat \rho_{A,B} \ket{1}$ and $\bra{2}\hat \rho_{A,B} \ket{2}$ correspond to the two-photon subtracted squeezed vacuum, the single-photon subtracted squeezed vacuum and the squeezed vacuum used in the protocol.}
\label{fig:fig7}
\end{figure*}

\subsection{Model of the generation scheme}
\label{sec:model_qutrit}
In order to implement a two-photon detection in the conditioning path, the approximation of the beam-splitter operator should be extended to the second order:
\begin{equation}
\begin{split}
\hat B(\theta) &= \text{e}^{\theta (\hat a \hat b^\dag-\hat a^\dag \hat b)} \\
&\approx 1+\theta (\hat a \hat b^\dag-\hat a^\dag \hat b) + \frac{\theta ^2 (\hat a \hat b ^\dag-\hat a^\dag \hat b)^2}{2}\ .
\end{split}
\end{equation}
Similarly, the two-mode squeezed vacuum state should also be written as:
\begin{equation}
\ket{\textrm{TMSS}} \propto \ket{0}_c \ket{0}_d + \lambda \ket{1}_c \ket{1}_d + \lambda^2 \ket{2}_c \ket{2}_d\ .
\end{equation}
Following the procedure detailed in section \ref{sec:hybrid1}, we can obtain the entangled state corresponding to a two-photon detection as
\begin{equation}
\frac{\lambda^2 r^2}{\sqrt 2}\ket{cat+}_a\ket{2}_d+ \theta \lambda tr\hat a \ket{cat+}_a\ket{1}_d+ \frac{\theta ^2 t^2}{2} \hat a^2 \ket{cat+}_a\ket{0}_d\ .
\end{equation}
In practical implementations, where the even cat state is approximated by a squeezed vacuum state, the above expression can be reformulated as
\begin{equation}
\label{eq:hybrid2sqstate}
\hat{S}_a(\zeta)\left(\mu^2 \ket{2}_d \ket{0}_a + \sqrt 2 \mu \ket{1}_d \ket{1}_a+ \ket{0}_d \ket{2}_a + c\ket{0}_d \ket{0}_a\right)
\end{equation}
where we use again equations \ref{eq:Sqsub1} and \ref{eq:Sqsub2} and the parameter $c = 1/(\sqrt 2 \tanh \zeta)$.

\subsection{Photon-counting balancing condition}
To maximize the achievable entanglement, balancing of Alice's and Bob's count rates should be obtained. We denote the coincidence counts with simultaneous detection of two photons from either Alice's or Bob's side as $C_{A,A}$ and $C_{B,B}$, respectively. We also denote with $N_A$ and $N_B$ the single-photon counts from each side in an acquisition time $T$. These counts are linked by the second-order auto-correlation functions $g_A$ and $g_B$ as
\begin{equation}
\begin{split}
 C_{A,A} &= g_A N_A^2 \tau/T \\
 C_{B,B} &= g_B N_B^2 \tau/T \ .
 \end{split}
\end{equation}
The auto-correlation functions $g_A$ and $g_B$ corresponding to the thermal state in Alice's side and the squeezed vacuum state on Bob's side are given by
\begin{equation}
 g_A = 2 \quad \textrm{and} \quad g_B = 3+\frac{1}{\sinh^2\zeta} \ .
\end{equation}
Therefore, balancing the coincidence counts $C_{A,A}$ and $C_{B,B}$ leads to the condition
\begin{equation}
\mu^4 = \frac{3}{2} + \frac{1}{2\sinh^2 \zeta} = 1 + c^2\ .
\end{equation}

\subsection{Representation of the hybrid entangled state}
The state presented in \ref{eq:hybrid2sqstate} can be reformulated as
\begin{equation}
\label{eq:hybrid2}
\begin{split}
\ket{\Psi}_{AB} = \hat{S}_B(\zeta) \big[\mu^2 \ket{2}_A \ket{0}_B + \sqrt 2 \mu \ket{1}_A \ket{1}_B+\\
 \ket{0}_A(c\ket{0}_B + \ket{2}_B)\big]/\sqrt{c^2+(1+\mu ^2)^2}\ ,
\end{split}
\end{equation}
where the subscripts $d, a$ are replaced by $A, B$. Figure \ref{fig:fig7}(a) displays the hybrid representation of the qutrit entangled state for the balancing condition $\mu^4 = 1 + c^2$. The diagonal blocks $\bra{0}\hat \rho_{A,B}\ket{0}$, $\bra{1}\hat \rho_{A,B} \ket{1}$ and $\bra{2}\hat \rho_{A,B} \ket{2}$ correspond to the two-photon subtracted squeezed vacuum, the single-photon subtracted squeezed vacuum and the squeezed vacuum, respectively. The two-photon subtracted squeezed vacuum is a good approximation to an even cat state. By using 6-dB local squeezing, the cat size in the generated hybrid entangled qutrit is enlarged with the presence of pronounced negativity in the Wigner function as shown in figure \ref{fig:fig7}(b) .

\begin{figure*}[t]
\centering
\includegraphics[width=0.85\textwidth]{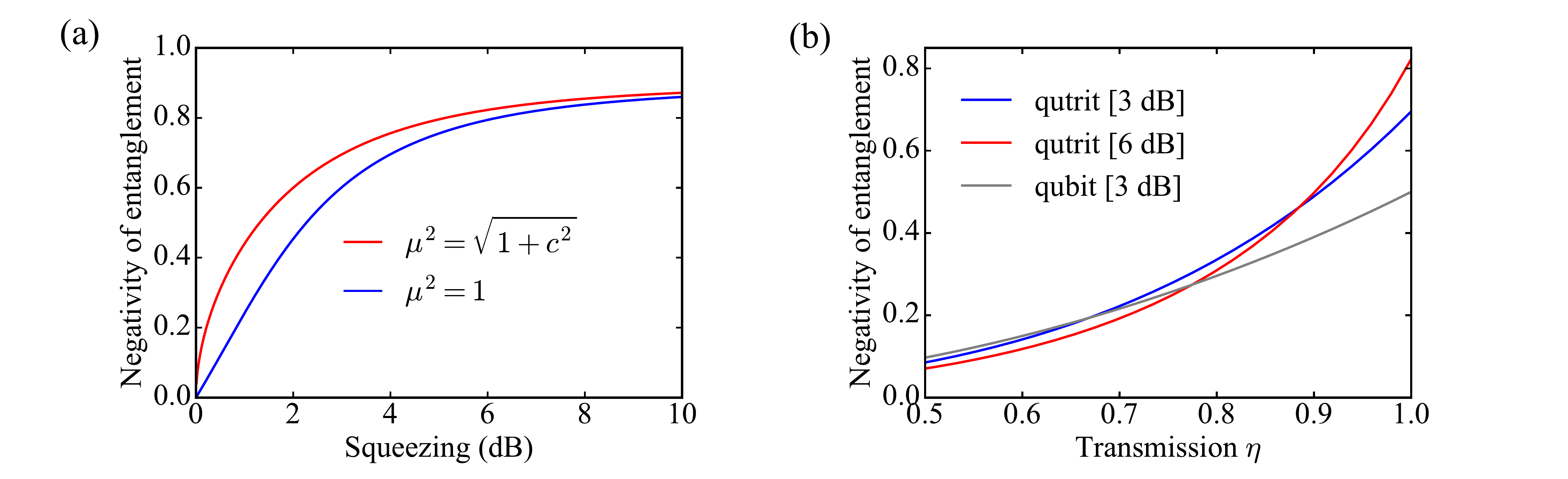}
\caption{(a) Negativity of entanglement for the hybrid qutrit state as a function of the local squeezing level in the single-photon and two-photon balancing conditions. (b) Negativity of entanglement for hybrid qutrit entanglement as a function of the transmission. For comparison, the negativity is also given for hybrid qubit entanglement in gray.}
\label{fig:fig8}
\end{figure*}

\subsection{Negativity of entanglement}
The negativity of entanglement for the hybrid qutrit state given in equation \ref{eq:hybrid2}, in the absence of losses, can be written as
\begin{widetext}
\begin{equation}
\mathcal{N} = \frac{\mu^2 + \mu \sqrt{1+c^2+\mu^4-\sqrt{c^4+2c^2(\mu^4+1)+(\mu^4-1)^2}} + \mu \sqrt{1+c^2+\mu^4+\sqrt{c^4+2c^2(\mu^4+1)+(\mu^4-1)^2}}} {c^2+(\mu^2+1)^2} \ .
\end{equation}
\end{widetext}
The maximum negativity $\mathcal{N}$ is ahieved under the balancing condition $\mu^4 = 1 + c^2$ and is given by
\begin{equation}
\begin{split}
\mathcal{N}_{\max}(c) = \sqrt {(1+c^2)} \bigg( \sqrt{2\sqrt{1+c^2}-2c} \ + \\\sqrt{2\sqrt{1+c^2}+2c} \ + 1 \bigg ) \bigg / \bigg[ 2(1+c^2+\sqrt{1+c^2})\bigg]\ .
\end{split}
\end{equation}
For a qutrit entangled state, the maximum achievable value for $\mathcal{N}$ is 1. In the limiting case of very large squeezing, i.e.\ $c \to 1/\sqrt{2}$, the negativity reaches here $\mathcal{N}_{\max}|_{c=1/\sqrt 2} \approx 0.9$. The state is therefore not maximally entangled.

As shown in figure \ref{fig:fig8}(a), the negativity of entanglement increases with larger squeezing values. With 6-dB initial squeezing, a negativity of 0.82 can be obtained. Therefore, even for moderate squeezing the hybrid entangled qutrit could exhibit a negativity of entanglement stronger than what could at any point be attained by hybrid entangled qubits.

We now study the negativity of entanglement in the presence of loss. Figure \ref{fig:fig8}(b) presents the value of $\mathcal{N}$ as a function of transmission. In general, local squeezing makes the decoherence faster due to the enlarged ``size'' of the quantum state. With a local squeezing of 6 dB, the entanglement of the hybrid qutrit state drops below the one obtained with 3-dB squeezing when the channel transmission is smaller than 88\%. Furthermore, the available hybrid qutrit entanglement drops below the one of the hybrid qubit state for a transmission below 77\%. Therefore, the potential higher entanglement for the hybrid qutrit states can only be observed in the condition of low overall loss, i.e. high-purity initial sources, high-transmission propagation channels and high-efficiency detection.

\section{Conclusion}
\label{sec:conclusion}
In conclusion, we have presented various schemes to engineer hybrid entanglement between CV and DV optical states. For each scheme we have identified the main parameters shaping the entangled state, such as the photon-counting balancing parameter and the local squeezing level in the experimental implementation. Moreover, we have investigated how these parameters determine important properties of the heralded state, such as the negativity of the Wigner function, the fidelity with the targeted states and the overall degree of entanglement. For the generation of hybrid entangled qubits we have shown in particular that a local single-photon subtraction in the CV mode leads to a higher fidelity with a cat size $|\alpha|^2$ greater than 1. The demonstrated size is compatible with the values $|\alpha|^2 \approx 1$ shown as the optimal value in recent proposals for resource-efficient operations with hybrid qubits \cite{Lee2013PRA}. Furthermore, we have analyzed a novel scheme to prepare hybrid qutrit entanglement by applying a two-photon heralding detection.

Experimentally, these states are challenging to prepare. They require non-classical states with high purity, and sometimes strong squeezing, together with large efficiencies for state propagation and detection. Thanks to recent progresses in optical quantum state engineering, these advanced schemes are becoming increasingly more feasible. The generation of these hybrid entangled states would provide crucial light sources to explore a variety of novel protocols for heterogeneous quantum networks.

\section*{Acknowledgments}
This work was supported by the European Research Council (Starting Grant HybridNet), Sorbonne Universit\'e (PERSU program) and the French National Research Agency (Hy-Light project). G.G. acknowledges the financial support from the EU (Marie Curie fellowship Helios) and T.D. from the Region Ile-de-France in the framework of DIM SIRTEQ. K.H. is supported by Program for Professor of Special Appointment (Eastern Scholar) at Shanghai Institutions of Higher Learning, Science and Technology Innovation Program of Basic Science Foundation of Shanghai (18JC1412000).

\end{document}